\documentclass[reprint,aps,prl,floatfix,amsmath,amssymb]{revtex4-2}
\usepackage{graphicx}
\usepackage{bm}

\begin{document}

\title[]{Electron heating and radiation in high aspect ratio sub-micron plasma generated by an ultrafast Bessel pulse within a solid dielectric}








\author{Kazem Ardaneh} 
\email{kazem.arrdaneh@gmail.com}
\affiliation{FEMTO-ST Institute, Univ. Franche-Comt\'e, CNRS, 15B avenue des Montboucons, 25030 Besan\c{c}on cedex, France.}
\affiliation{Sorbonne University, Pierre and Marie Curie Campus, 4 place Jussieu, 75252 Paris Cedex 5, France}

\author{Remo Giust, Pierre-Jean Charpin, Benoit Morel and Francois Courvoisier}
\email{francois.courvoisier@femto-st.fr}
\affiliation{FEMTO-ST Institute, Univ. Franche-Comt\'e, CNRS, 15B avenue des Montboucons, 25030 Besan\c{c}on cedex, France.}

\begin{abstract}
\vspace{1cm}
This preprint has not undergone peer review. The Version of Record of this article is published in The European Physical Journal Special Topics, and is available online at https://doi.org/10.1140/epjs/s11734-022-00751-y

\vspace{0.5 cm}

Full reference:  K. Ardaneh, R. Giust, P.-J. Charpin, B. Morel and F. Courvoisier " Electron heating and radiation in high aspect ratio sub-micron plasma generated by an ultrafast Bessel pulse within a solid dielectric ", {\it The European Physical Journal Special Topics}, (2022). DOI: 10.1140/epjs/s11734-022-00751-y

\vspace{1cm}

When propagating inside dielectrics, an ultrafast Bessel beam creates a high aspect-ratio cylinder of plasma with nanometric diameter that extends over several tens of micrometers to centimeters. We analyze the interaction between the intense ultrafast laser pulse and the plasma rod using particle-in-cell simulations. We show that electrons are heated and accelerated up to keV energies via transit acceleration inside the resonance lobes in the vicinity of the critical surface and compute their radiation pattern.
\end{abstract}

\maketitle

\section{Introduction}\label{sec1}

Ultrafast lasers are ideal tools to deposit energy within the bulk of transparent materials \cite{Gattass2008}. This has applications for laser micromachining, for the generation of new material phases as well as for the generation of warm dense matter. Thanks to the nonlinear ionization, the infrared radiation of the laser can generate, early in the pulse, a plasma of electrons and holes in the bulk of transparent dielectrics \cite{Rethfeld2017}. The interaction of the trailing part of the laser pulse can heat the plasma if proper conditions are met. Then, depending on the energy density that has been deposited within the plasma, phase change can occur, even at sub-picosecond time scale via non-thermal melting  if the ionization rate of the solid is sufficiently high (only few \%) \cite{SokolowskiTinten1998, Sundaram2002}. In this case, the material can quickly reach the warm dense matter regime, in which the material is at solid density but with temperatures on the order of 1-10~eV \cite{Engelhorn2015, Falk2018} . This is a challenging state still under study which  is relevant for the modeling of many astrophysical objects \cite{Guillot1999}. The phase change triggers also a series of physical phenomena in the material such as shockwave emission, void formation \cite{Glezer1997}, densification around the void \cite{Juodkazis2006}. In the densified regions around voids formed within the bulk, the extreme temperatures and pressures reached within a short time can lead to the formation of new material phases as it has been observed in sapphire and silicon \cite{Vailionis2011,Smillie2020}. Finally, the formation of voids inside the material has useful applications for laser cutting or drilling of transparent materials. High-speed (typ. 1 m/s) cutting of glass with the stealth dicing technology is one of the most relevant examples \cite{Mishchik2016,Meyer2017,Jenne2018a,Meyer2019}.

In these three domains of applications, it is clear that a challenge is to reach the largest energy density as possible over the largest volume possible. However, it is well-known that nonlinear filamentation of Gaussian beams prevents reaching extreme energy densities inside dielectrics. In contrast, we recently demonstrated that it is possible with Bessel beams \cite{Ardaneh2021}.

Zeroth-order Bessel beams, also called "diffraction-free" beams, constitute a propagation-invariant solution to the wave equation \cite{Durnin1987}. They are featured by a conical flow of light directed toward the optical axis. The conical interaction creates an interference pattern, characterized by an intense central lobe surrounded by several other circular lobes of lower intensity. Importantly, when propagating inside transparent solids, an ultrafast laser pulse shaped as a Bessel beam can generate a nano-plasma rod with a length that is adjustable independently of the diameter. Recent work has shown that the length of this plasma rod can be scaled from tens of micrometers to 1~cm \cite{Meyer2019}.

We have recently demonstrated that 100 fs Bessel beams can generate elongated plasma rods with over critical plasma density in sapphire $Al_2 O_3$ \cite{Ardaneh2021}, in the regime corresponding to the formation of high aspect ratio nano-voids \cite{Rapp2016}. 
In contrast with the case of the Gaussian beam, all the pulse energy in the Bessel beam impinges with a relatively large incidence angle toward the plasma rod generated early in the pulse along the optical axis. This configuration is ideal to trigger resonance absorption. In reference \cite{Ardaneh2021}, we have demonstrated the occurrence of resonance absorption inside sapphire by comparing experimental results to particle-in-cell (PIC) simulations. The strong energy transfer between the laser wave to the core of the plasma opens the possibility to reach the warm dense matter regime and to produce temperatures on the order of 10~eV. This explains the opening of high aspect ratio nano channels inside several dielectrics upon Bessel beam femtosecond illumination. Importantly, the conical geometry of Bessel beams is invariant along the propagation. This implies that results obtained with Bessel beams of only several tens of micrometers in length can be extrapolated to several centimeters.

A key question is the understanding of the microphysics of the interaction of the Bessel beam with the sub-micron plasma, at moderate intensities (typically $10^{14}$ W/cm$^2$). For this, we have used Particle-In-Cell (PIC) simulations to investigate the electron plasma wave generation, particle heating and acceleration, as well as the radiation emitted by the accelerated particles.

\begin{figure*} [h!]
    \centering
    \includegraphics[width=0.95\textwidth]{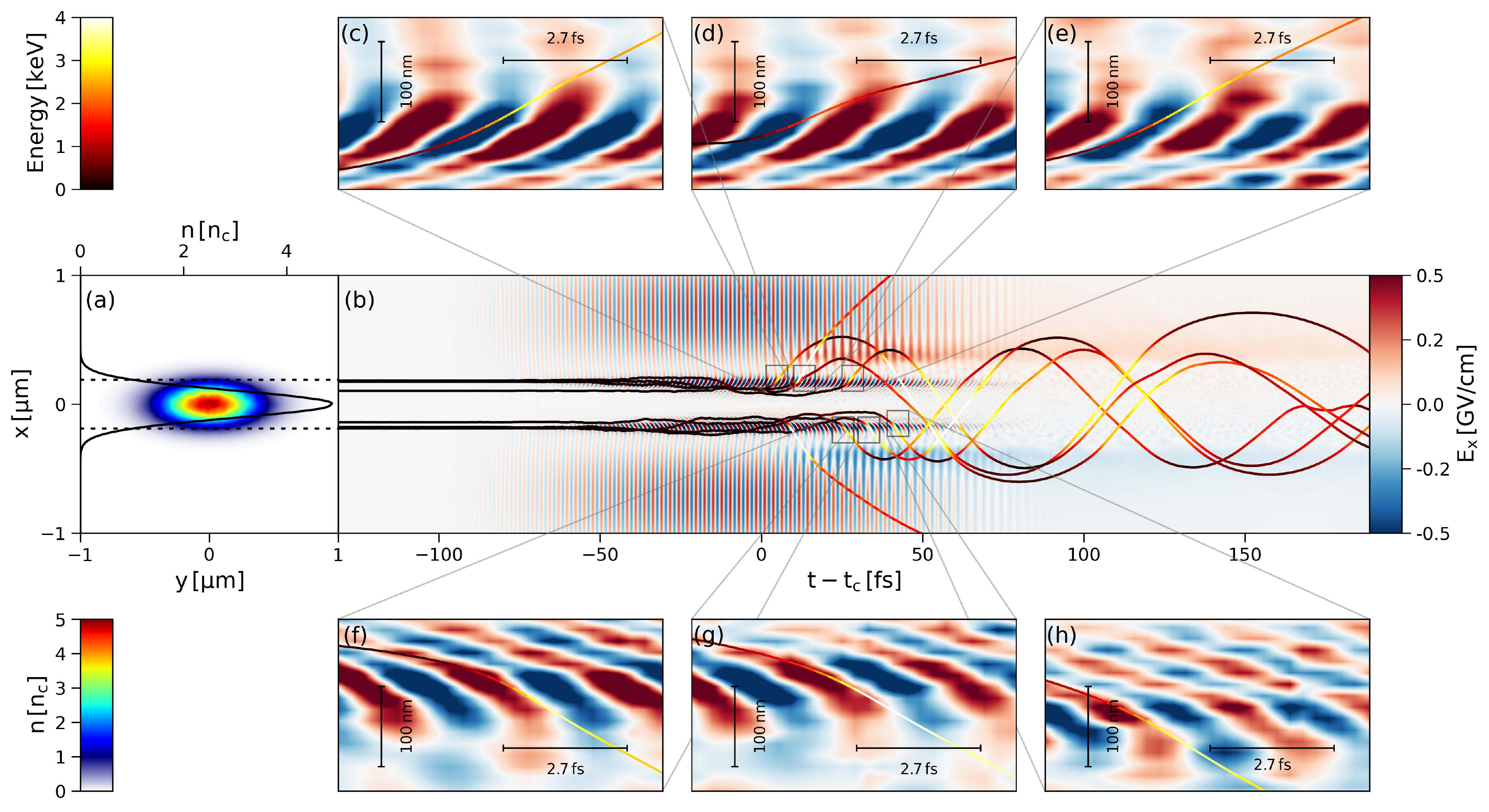}
    \caption{(a) Cross section of the 2D density profile of the plasma. The black solid line shows the density profile at $y=0$ along x-direction. The critical density $n_\mathrm{c}$ is $1.7 \times 10^{21}$~cm$^{-3}$ (b) $E_\mathrm{x}$ component of the electric field as a function of time and $x$ position. The solid colored lines correspond to the projection in the plane of 8 particle trajectories. Their energy is indicated using the color code. (c-h) Zoom-in views of the acceleration of the electrons as they leave the plasma, together with the exact value of the field $E_\mathrm{x} (x,y=y_p,z=z_p,t)$ sampled at the particle position ($x_p$,$y_p$,$z_p$, t).}
    \label{fig:Field_trajectories}
\end{figure*}

\section{Simulations}\label{sec:Simulations}
Our PIC simulations are based on the interaction of a Bessel-Gauss beam \cite{Ardaneh2020} with a preformed plasma. The 100~fs laser pulse, with a central wavelength of 800~nm, is polarized along $x$-direction. We assume that nonlinear ionization has produced early in the pulse a plasma and we model the interaction of the most intense part of the laser pulse with this pre-plasma. We used the numerical code EPOCH \cite{Arber2015}, without ionization, and the numerical scheme is detailed in reference \cite{Ardaneh2022}.
We maintained the numerical heating at a negligible level over the duration of the simulation (320~fs). The pre-plasma is a plasma rod extending over the whole longitudinal length of the box (because of the invariance of the Bessel beam), and its transverse cross-section is elliptically-shaped. The profile is Gaussian and is the same as the one matching our experimental results, as shown in reference \cite{Ardaneh2021}: the critical radius is 190~nm  along the polarization direction ($x$ axis)  and 380~nm in the other direction, as shown in Fig. 1(a). The peak intensity of the pulse is 6×10$^{14}$W/cm$^2$, for a pulse duration of 100~fs.

\section{Field amplification and particle acceleration}\label{sec:amplification}

Figure \ref{fig:Field_trajectories}(b) shows the evolution of the $E_x$ component of the electric field in time, on a segment placed along the $x$ direction at the propagation distance corresponding to the highest intensity reached in the Bessel-Gauss beam \cite{Boucher2018}. We first observe the field enhancement in the region where the permittivity decreases because of the presence of plasma ( $\vert  x\vert<190$~nm). A strong field amplification occurs on the critical surfaces, that are indicated with dashed lines in Fig. \ref{fig:Field_trajectories}(a). The field reaches a maximum value of 1~GV/cm. This corresponds to an amplification factor of approximately 7 in comparison with the input laser field.

\begin{figure*}
    \centering
    \includegraphics[width=0.9\textwidth]{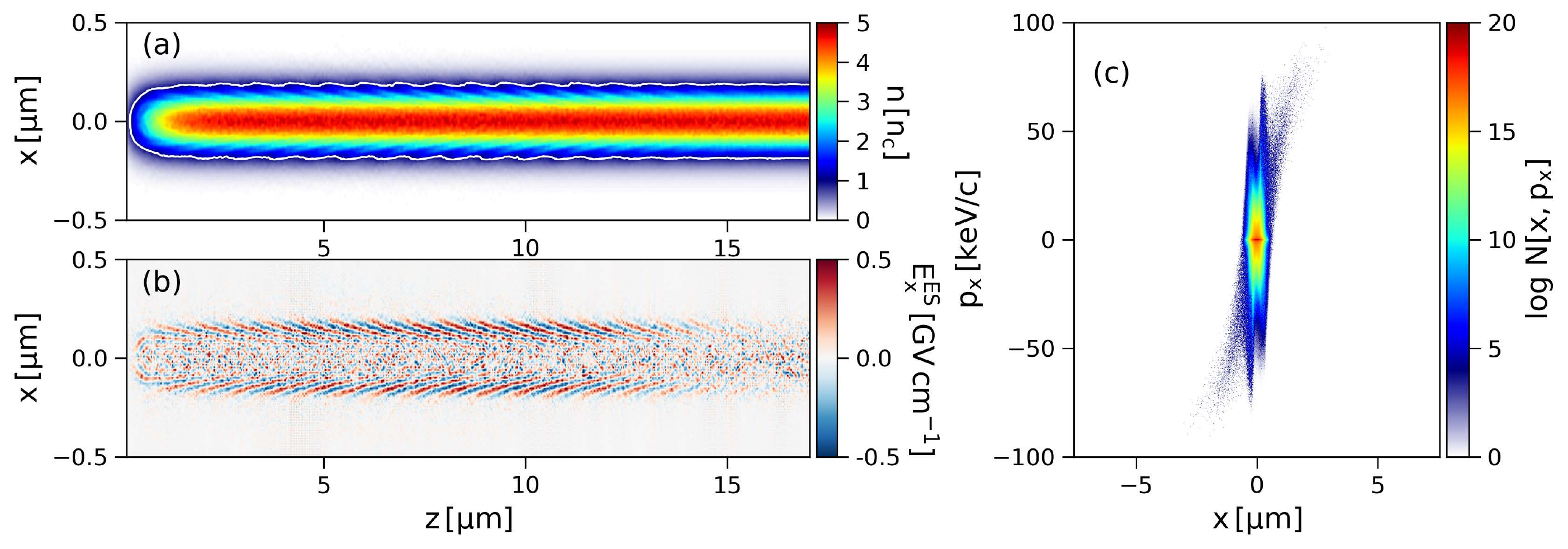}
    \caption{(a) Plasma density at the time corresponding to the peak intensity in the plane $y=0$, which is parallel to the polarization direction. The white solid line shows the contour at the critical density(b) Longitudinal component of the electric field, which allows the visualization of the plasma density waves. (c) $x-p_\mathrm{x}$ phase space of the electron population at a time 135~fs after the peak of the pulse.}
    \label{fig:Density_EPW_heating}
\end{figure*}

At the critical surface, the resonance absorption takes place. The wave conversion phenomenon creates electron plasma waves. We plot in Fig. \ref{fig:Density_EPW_heating}(a) the density of electrons at the peak of the pulse, in $x-z$ plane. The white line shows the contour at the critical density. We see plasma oscillations taking place. To allow a clearer visualization of the plasma waves, we computed the electrostatic field, that is shown in Fig. \ref{fig:Density_EPW_heating}(b). It has been derived from the density $\rho$  using Gauss's law in the Fourier $\mathbf{k}$-space, $E^{\rm ES}_{\rm x}=-j{\mathbf{k}}\rho({\mathbf{k}})/k^2/\epsilon_0$.

We can observe that, in the sub- critical region, outside the plasma, the electron plasma waves are quickly damped. This arises from the efficient Landau damping. In the over-critical region, the field oscillations are due to electron sound waves \cite{Ardaneh2022,Holloway_1991}. They penetrate relatively deeply into the overcritical region and we observe that they are curved. The propagation of these waves into the overcritical region is attributed to the fact that the temperature is highly inhomogeneous in the plasma. This is also the reason of the variation of the spatial period along $x$, hence the variation of the apparent curvature of the fringes. This structure will have a strong impact on the acceleration of electrons as we will see below.

The heating of the particles has been investigated in Ref. \cite{Ardaneh2021}. In summary, the wave particle energy exchange is very efficient: we observe the heating of the electron population mainly around the critical surfaces. In Fig. \ref{fig:Density_EPW_heating}(c), we show the particle distribution in the phase space $x-p_x$ after the pulse (135~fs). We evaluated the main component temperature to be 1.3~eV, and a hot electrons component at 70~eV. In addition, we observe in Fig. \ref{fig:Density_EPW_heating}(c), two tails expanding outwards, at high momentum, outside the plasma.

These tails correspond to highly accelerated particles with energies up to  7~keV.
Figure \ref{fig:Field_trajectories} provides a glimpse on the physical mechanism of the acceleration. We have selected out 1000 of the most energetic particles and have traced their trajectories. In Fig. \ref{fig:Field_trajectories}(b), we have plotted a selection of 8 representative trajectories. Two of them show highly accelerated particles that escape from the plasma. The other 6 are accelerated by the same mechanism, but remain trapped by a static electric field generated by a double layer formation that will be explained later. The acceleration mechanism is the same in all cases: in the different sub-figures \ref{fig:Field_trajectories}(c-h), we see that the particles undergo transit acceleration, which occurs when a particle travels through a highly non-uniform electromagnetic field \cite{Shen1965,Morales1974,DeNeef1977}. The electrons gain energy by riding on the electron plasma wave, that is curved in the $x-t$ space. The effective acceleration occurs on a distance of less than 60~nm, when crossing the high resonance field. Depending on the exact position of the particle with respect to the plasma wave, the acceleration is more or less efficient. We see that the curvature of the plasma wave is a key to obtain a progressive acceleration of the particle while it remains on the peak of the wave.

Particle acceleration and heating transfer a fraction of the electrons away from the main plasma, as it is also apparent on the $x-p_x$ representation of Fig. \ref{fig:Density_EPW_heating}(c). This generates a so-called double layer on either sides of the plasma. These double layers are apparent in Fig. \ref{fig:Field_trajectories} because they generate a static E-field that is superimposed to the laser pulse. Importantly, this static field has an amplitude that is as high as the resonance field of the laser pulse itself (on the order of GV/cm). This static E-field even remains after the laser pulse has vanished. Its damping is governed by the collisions inside the plasma.

\begin{figure}
    \centering
    \includegraphics[width=\columnwidth]{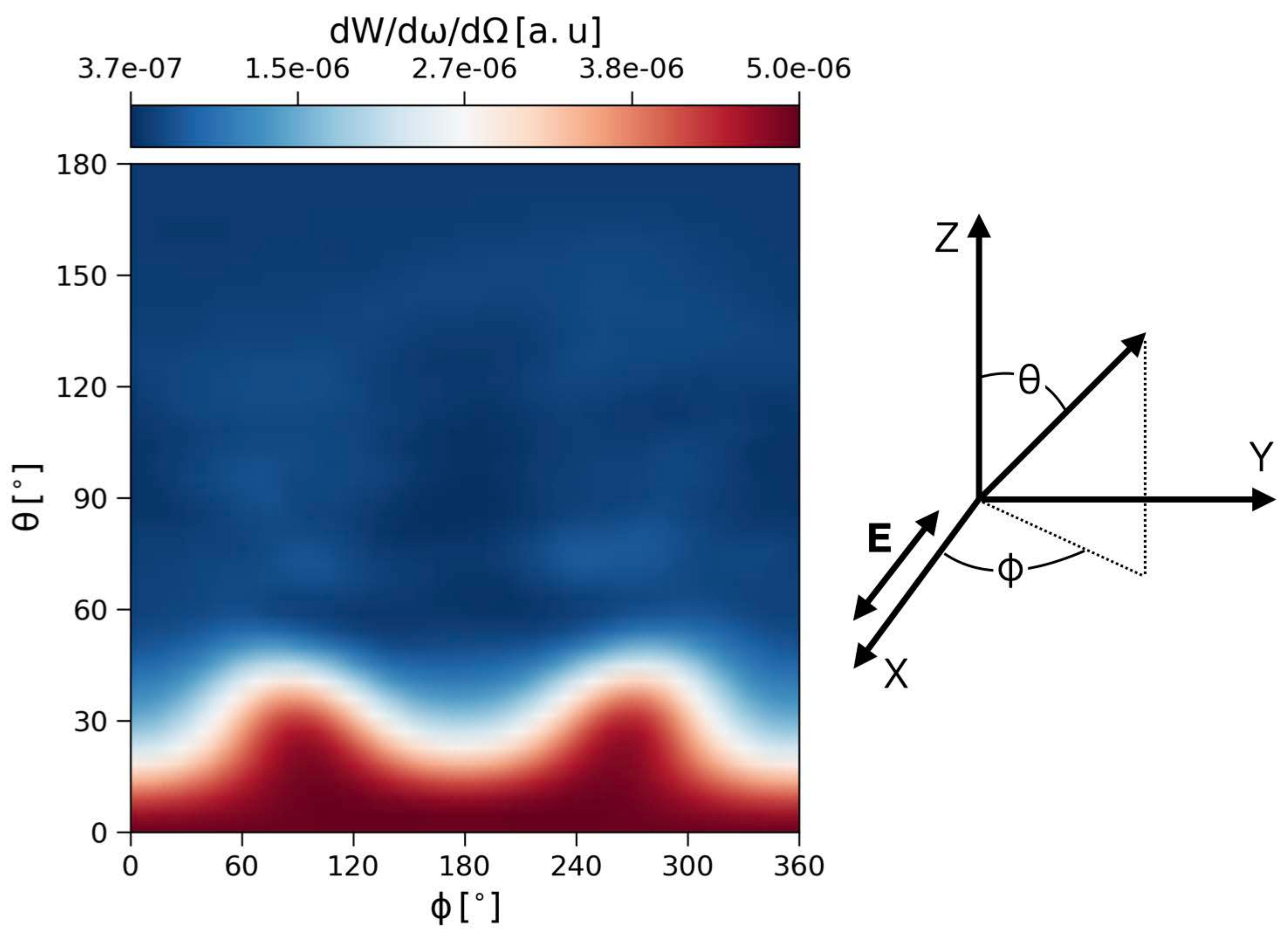}
    \caption{ (left)The total energy radiated per unit solid angle per unit frequency
from the accelerated electrons. (right) Configuration of the angles. The laser polarization is along the $x$ axis}
    \label{pow_spe_fig}
\end{figure}

\section{Radiation pattern}\label{sec:Radiation}

One of the major advantages of PIC codes is the possibility to access the full information about the particle dynamics, e.g., the position and the momentum as a function of time. If this information can be retrieved and stored for a number of particles, it is then feasible to post-process the radiation associated with a particular set of particles.  The radiation diagnostic uses the information from the particle trajectories, position and momentum over time, and determines the energy being radiated by an accelerated charged particle. 

Let us consider a particle at position ${\bf r}_{0}\left(t\right)$ at time $t$. At the same time, we observe the radiated electromagnetic fields from the particle at position ${\bf r}$.  Due to the finite velocity of light, we observe the particle at an earlier position ${\bf r}_{0}\left(t^{\prime}\right)$ where it was at the retarded time $t^{\prime}=t-{\bf R}\left(t^{\prime}\right)/ c$,
where $\mathbf{R}\left(t^{\prime}\right)=\vert\mathbf{r}-\mathbf{r}_{0}\left(t^{\prime}\right)\vert$ is the distance from the charged particle (at the retarded time $t^{\prime}$ ) to the observer. Using the Li\'enard-Wiechert potentials, the total energy $W$ radiated per unit solid angle per unit frequency from a charged particle moving with instantaneous velocity $\bm{\beta}=\mathbf{v}/c$ under acceleration $\dot{\bm{\beta}}=\mathbf{a}/c$ can be expressed as \cite{jackson_1998}:

\begin{equation}\label{power_omega}
\frac{{\rm d}^{2} W}{{\rm d} \omega {\rm d} \Omega}\propto\left\vert\int_{-\infty}^{\infty} \frac{\hat{\mathbf{n}}\times[(\hat{\mathbf{n}}-\bm{\beta}) \times \dot{\bm{\beta}}]}{(1-\bm{\beta} \cdot \hat{\mathbf{n}})^{2}} e^{i \omega(t-\hat{\mathbf{n}} \cdot \mathbf{r}(t) / c)} d t\right\vert^{2}
\end{equation}

Here, $\mathbf{n}=\mathbf{R}\left(t^{\prime}\right) /\vert\mathbf{R}\left(t^{\prime}\right)\vert$ is a unit vector that points from the particle retarded position towards the observer. The observer's viewing angle is set by the choice of $\mathbf{n}\left(\hat{\bf x}\sin\theta\cos\phi+\hat{\bf y}\sin\theta\sin\phi+\hat{\bf z}\cos\theta\right)$.

Figure \ref{pow_spe_fig} shows the total radiated energy from an ensemble of 100 randomly selected electrons traced in the simulations that we computed using Eq. (\ref{power_omega}). The distribution is plotted as a function of the angles $(\theta, \phi)$ in Fig. \ref{pow_spe_fig}(left). The forward direction corresponds to values of $\theta$ below 90$^{\circ}$.  We observe that the forward signal shows maxima at $(\theta, \phi)\approx(0, \pi/2)$ and $(0, 3\pi/2)$, perpendicular to the electron acceleration in the $x-$direction, which indeed corresponds to the pump laser polarization direction. The emission follows the well-known power distribution per solid angle $\Omega$ emitted by a single particle  \cite{jackson_1998}:

\begin{equation}\label{power_spe}
\frac{d P}{d\Omega}\propto \frac{\vert \dot{\bm{\beta}} \vert^{2}}{(1-\beta \cos \theta)^{3}}\left[1-\frac{\sin ^{2} \theta \cos ^{2} \phi}{\gamma^{2}(1-\beta \cos \theta)^{2}}\right]
\end{equation}

\noindent where $\gamma$ is the Lorentz factor.

Noticeably, the power shows a much higher signal in the angles corresponding to the forward direction than for the backward direction ($90^{\circ}<\theta\leq180^{\circ}$). This behaviour could be explained by the coherence of the phases of the  dipole moments induced along the plasma rod.
A more detailed analysis of the radiation spectrum, out of the scope of the present article, shows the presence of second harmonic generation and of THz radiation in the forward direction.

In conclusion, we have investigated the interaction between a moderately intense laser pulse shaped as a Bessel beam with a nanoplasma rod using particle-in-cell simulations. We have demonstrated that resonance absorption generate plasma waves inside plasma. These plasma waves are highly damped in the sub- critical region while plasma sound waves can propagate over several hundreds of nanometers inside the overcritical plasma. The analysis of the trajectories of the most energetic particles shows that the main acceleration mechanism is transit acceleration. It occurs at the critical layer when particles are trapped inside a plasma wave and gain energy until they are released at the critical surface. Because of the heating of the electron gas, two double layers form on either sides of the nano-plasma. The most energetic particles, with energies up to 7 keV can escape the plasma while part of the hot electrons remain trapped by the potential well due to the electrostatic field of the double layer. Overall, our results enable us to gain insights into the micro physics of the laser-plasma interaction that this relevant for the understanding of the different mechanisms of the deposition of the femtosecond laser pulse energy inside dielectrics. Our results reveal a rich physics which can be exploited in several fields of applications: laser-matter interaction, laser micro-machining, warm dense matter and high energy density physics inside solids, as well as the generation of electrostatic fields and terahertz radiation.

Acknowledgments :
The research leading to these results has received funding from the European Research Council (ERC) under the European Union's Horizon 2020 research and innovation program (grant agreement No 682032-PULSAR), R\'egion Bourgogne Franche-Comt\'e, I-SITE BFC project (contract ANR-15-IDEX-0003), and the EIPHI Graduate School (ANR-17-EURE-0002). We acknowledge the support of PRACE HPC resources under the Project "PULSARPIC" (PRA19\_4980 and RA5614), and GENCI resources under projects A0070511001 and A0090511001.

Data availability statement: Data will be made available on reasonable request.


\bibliography{main}


\end{document}